\input amstex.tex
\input amsppt.sty
\documentstyle{amsppt}
\hcorrection{15 true mm}
\vcorrection{20 true mm}
\pagewidth{124 true mm}
\pageheight{186 true mm}

\define\cl#1{\overline{#1}}

\topmatter

\title
Testing QM : the Uncertainty Principle 
\endtitle

\author
Ji\v r\'\i \space Sou\v cek 
\endauthor

\address
Faculty of Mathematics and Physics,
Charles University in Prague\linebreak
Sokolovsk\'a 83, 186 00  Praha 8,
Czech Republic
\endaddress

\email
soucekj\@karlin.mff.cuni.cz
\endemail

\abstract
We propose the experimental test of the uncertainty principle. From sub-quantum models it follows that the uncertainty principle may be not true on short time intervals of the order of a picosecond. The positive result of this experiment would signify the limits of QM. 
\endabstract


\thanks
This research was supported by grant No. RN19982003014 of the Ministry of Education.
\endthanks

\endtopmatter


\document
\loadeurm
\loadbold

In this paper we propose the experimental test of Heisenberg`s Uncertainty principle in Quantum Mechanics (QM). We propose that the "physical" uncertainty principle may be not true on very short intervals of time. 

The idea of such a proposal originates from the sub-quantum models introduced in [1,2]. The phenomenon considered here was suggested in [1] and introduced as a "gedanken" experiment in [2] under the name "concentration effect".

The phenomenon predicted here depends, as subquantum models in general, on a certain constant $\tau_0$ with the dimension of time and with the meaning of the relaxation time. On time intervals much greater then $\tau_0$ the behavior of the subquantum model approaches the standard QM behavior.

In the subquantum model (SubQM) it is assumed that particles move deterministically (but with a probability amplitude) under certain random force (Einstein's position, but with a probability amplitude distribution for this random force) and the constant $\tau_0$ is related to the strength of this random force. It is assumed that this random force comes from the interactin with the cosmological dark energy and this relation enables us to estimate the value of $\tau_0$. The resulting estimation gives (very roughly)
$$
\tau_0 \gtrsim 1 ps \quad(picosecond) \thickspace .
$$

The proposed experiment consists in the diffraction of the light going through the repeated single slit and then being observed at the screen. All passage of fotons through the instrument should be as short as possible, at least of the order of $\tau_0$.

The quantity measured in this experiment is the spreading out (the dispersion) $\Delta x$ of the probability distribution of observed fotons on the screen, and namely, the dependence of $\Delta x$ on the distance $L$ between the slits. The geometry of the proposed experiment is described in the Fig. 1.

So that we have $\Delta x \thickapprox 4 \cdot x_{1null}$, where $x_{1null}$ denotes the first zero (the first minimum) of the probability density of photons at the screen. We shall assume that the width of the slit $\delta_0$ is relatively small with respect to the distance between the second slit and the screen $L_0$ and that $L_0$ is smaller then $L$
$$
\delta_0 \ll L_0 \quad, \quad L_0 \le L \quad.
$$
If $\lambda_0$ is the wave length of the light used in the experiment, then we have relations
$$
x_{1null}\cong L_0 \cdot \frac{\lambda_0}{\delta_0} \quad,\quad x_{k\thinspace null}=k \cdot x_{1null} \quad.
$$

This follows from the Fig.2.

We have in general
$$
\Delta x \approx 4 \cdot x_{1null} = 4 \cdot x_{1null}(\delta_0,\lambda_0,L_0,L) \quad .
$$

In QM we know, that $\Delta x$ (essentially) does not depend on $L$ ,$$ d\Delta x /dL \approx 0 . $$ Our prediction (from SubQM) says that $\Delta x$ {\bf does depend} on $L$ $$ d\Delta x /dL > 0 \quad,  $$ for example in the way exemplified in the Fig.3.

The critical length $L_{crit}$ is related to $\tau_0$ by
$$
L_{crit} \approx c \cdot \tau_0 \quad (c= velocity \thickspace of \quad light)
$$
and we have to assume that $$L_{crit} \gtrsim L_0 \quad .$$ 

This means that in SubQM the dispersion of the light $\Delta x$ is smaller than in QM, provided $L \lesssim L_{crit}$. So that we have for short time our prediction
$$
\Delta x^{(SubQM)} < \Delta x^{(QM)} \quad for \quad L \lesssim L_{crit} \quad ,
$$
while for long times we have  the standard QM result
$$
 \Delta x^{(SubQM)} = \Delta x^{(QM)} \quad for \quad L \gg L_{crit} \quad .
$$ 

Using the paper [3] it can be found that the following values of parameters are experimentally realizable
$$
\align
\lambda_0 &\cong 700 \thinspace nm = 0.7 \cdot 10^{-3} \thinspace mm \thickspace,\\
\delta_0 &\cong 10 \thinspace \mu m = 0.01 \thinspace mm \thickspace,\\
L_0 &\cong 0.3 \thinspace mm \thickspace.
\endalign
$$

Then we obtain
$$
\Delta x \thickspace \cong \thickspace 4 \cdot L_0 \cdot \frac{\lambda_0}{\delta_0} 
\thickspace \cong \thickspace 4 \thickspace \cdot 0.3 \cdot \frac{0.7 \cdot 10^{-3}}{10 \cdot 10^{-3}}
\thickspace \cong \thickspace 0.1 \thinspace mm
$$
so that the necessary relations
$$
L_0 \lesssim c \cdot \tau_0 \thickspace , \thickspace 
\Delta x \gg \delta_0 \thickspace , \thickspace 
L_0 \gg \delta_0 
$$
are satisfied.

\hphantom {-}
The proposed experiment requires to proceed through the following steps : \newline
1. To change the distance $L$ between slits from $L_0 \approx 0.3 \thinspace mm $ to $ 0.3 \thinspace meter$ \newline
2. To measure $\Delta x$ for different values of $L$ holding other parameters ($\lambda_0, \delta_0, L_0$) fixed \newline
3. To compare $\Delta x$'s for different $L$'s and test in this way our prediction - QM requires that $\Delta x$ does not depend on $L$, while the non-trivial depedence of $\Delta x$ on $L$ indicates the SubQM behavior.

\hphantom {-}
The value $ \Delta x $ can be defined in the mathematically precise way as follows. Let $p(x)\,dx$ be the {it normalized} density of fotons arrived (during the unit of time) at the interval $[x,x+dx]$, i.e. $\int p(x)\,dx = 1$. Then the diameter $\Delta x$ of the dispersion of the light on the screen can be, for example, defined by
$$
\Delta x = 2 \cdot min \big\{R>0 : \thickspace exists \thickspace x_0 \thickspace such \thickspace that \thickspace 
\int_{x_0 - R}^{x_0 + R}p(x)\,dx \ge 0.7 \big\}.
$$

\quad

The estimation of $\tau_0$ in SubQM is based on the following considerations :\newline
(i) it is assumed that quantum particles move deterministically (this "deterministical" QM is introduced in [2]), but under the influence of a certain random force \newline
(ii) this random force is represented as a result of the interaction of a particle with certain medium \newline
(iii) it is supposed that this medium is the cosmological dark energy represented as a system of particles moving with the velocity $\ge c$ (tachyons) \newline
(iv) the estimate of the mean density of normal particles is of the order \newline $\phantom{xxxxxxx} 1 \thickspace baryon / m^3$ \newline and let us assume that the density of all normal particles are of the order \newline $\phantom{xxxxxxx} 100 \thickspace particles / m^3$ \newline
(v) using the fact that the density of the dark energy is $\approx$ 10 times larger than the density of normal particles we arrive at the estimate \newline 
$ 
\phantom{xxxxxxx} 1000 \thickspace "dark \thickspace energy \thickspace particles" / \thickspace 3 \thickspace nanoseconds
$ \newline since $c \cdot 3 \thickspace ns \approx 1 \thickspace m$ \newline
(vi) thus the interaction of a "deterministical" sub-quantum particle with the dark energy particles goes with the mean frequency \newline 
$
\phantom{xxxxxxx} 1 \thickspace interaction \thickspace for \thickspace each \thickspace 3 \thickspace ps \thickspace (picosecond)
$ \newline
(vii) thus we can reasonably estimate the relaxation time by \newline 
$
\phantom{xxxxxxx} \tau_0 \gtrsim 1 \thickspace ps \qquad(and \thickspace perhaps \thickspace \tau_0 \gg 1 \thickspace ps).
$ \newline
Then we have \newline 
$
\phantom{xxxxxxx} L_{crit} \approx c \cdot \tau_0 \gtrsim 0.3 \thickspace mm \thickspace .
$

\hphantom{-}
The mechanism of SubQM, which creates the behavior described above can be summarized by : \newline
(i) in SubQM the wave function depends not only on positions of particles (as in QM), but also on the velocities of particles \newline
(ii) after the free evolution during the time interval $\Delta t \gg \tau_0$, all possible velocities of a given particle are equally probable (the mean value of the absolute value of velocity is infinite)\newline
(iii) during time intervals $\Delta t \ll \tau_0$, the velocity of a particle changes (under the influence of a random force), but not very much \newline
(iv) if particle passed from the point $A$ to another point $B$ during the time interval $ \Delta t \ll \tau_0$, then its velocity at $B$ is close to the velocity of a classical motion between these two points, so that the selection by our two slits creates the concentration of velocities \newline
(v) if particle passed from the point $A$ to another point $B$ during the time interval $\Delta t \gg  \tau_0$, then its velocity at $B$ is more or less arbitrary \newline
(vi) thus if $\Delta t \ll \tau_0$, then possible velocities of particles are concentrated around the classical velocity, while if $\Delta t \gg \tau_0$, then possible velocities of particles are not concentrated, but distributed along all space \newline
(vii) if velocities at the second slit are concentrated and $L_0 \le L_{crit}$ holds, we obtain that the dispersion of particles on the screen will be much smaller than that predicted by QM, because the velocity of a particle (selected by slits) is always closed to the classical velocity \newline
(viii) the phenomenon of the relaxation (for $\Delta t \gg \tau_0$) consists in the fact that the distribution of velocities in the subquantum wave function is such that all velocities are almost equally probable (on the other hand, the selection done by the two consecutive slits creates the concentration of possible velocities such that then the uncertainty principle does not hold). \newline
All these arguments can be found (together with necessary calculations) in [2].

\quad


\enddocument
\bye